\documentclass[a4paper,11pt]{article}
\pdfoutput=1
\usepackage{jheppub}
\usepackage{terms} 

\usepackage{amsmath}
\usepackage{amssymb}
\usepackage{physics}
\usepackage{graphicx}
\usepackage{hyperref}  
\hypersetup{
	colorlinks = true,
	linkcolor = blue,
	citecolor = blue,
	urlcolor = blue
}
\usepackage{tensor}
\usepackage[normalem]{ulem}

\newcommand{\MP}{M_\mathrm{P}}

\title{\boldmath Phenomenology of wavelike vector dark matter nonminimally coupled to gravity}

\author{Hong-Yi Zhang}
\author{and Siyang Ling}
\affiliation{Department of Physics and Astronomy, Rice University, Houston, Texas 77005, USA}

\emailAdd{hongyi@rice.edu}
\emailAdd{siyang.ling@rice.edu}

\abstract{We study three astrophysical/cosmological consequences of nonminimal couplings to gravity in wavelike vector dark matter. In the nonrelativistic limit, the nonminimal coupling with the lowest mass dimension leads to effective self-interactions that affect the mass-radius relation of vector solitons, growth of linear perturbations during structure formation, and the speed of gravitational waves (GWs). Based on the success of cold dark matter on large-scale perturbations and the current limits on GW speed, we constrain the dark matter mass and nonminimal coupling strength to be within the range $\abs{\xi_1} / m^2 \ll 10^{50} \mathrm{eV^{-2}}$ and $-3\times 10^{46} \mathrm{eV^{-2}} \lesssim \xi_2 / m^2 \lesssim 8 \times 10^{48} \mathrm{eV^{-2}}$.}

\begin{document} 
\maketitle
\flushbottom

\section{Introduction}
\label{sec:intro}

The nature of dark matter (DM) continues to be a central objective of modern cosmology. Among various types of dark matter models (see \cite{Bertone:2016nfn, Roszkowski:2017nbc, Bernal:2017kxu, Schumann:2019eaa, Lin:2019uvt, ParticleDataGroup:2022pth} for reviews), vector dark matter (VDM) has been drawing increasing attention of theorists and experimentalists in recent years \cite{Fabbrichesi:2020wbt, Caputo:2021eaa}. Proposals for VDM cover a large range of mass, with a division occurring at $30\mathrm{eV}$ \cite{Hui:2021tkt}, below or over which the vector bosons behave like waves \cite{Freitas:2021cfi, Adshead:2021kvl, Gorghetto:2022sue, Amin:2022pzv, Antypas:2022asj} or particles \cite{Lebedev:2011iq, Djouadi:2011aa, Baek:2012se, Arcadi:2020jqf, Baouche:2021wwa, Ghorbani:2021yiw}. VDM can be produced through, for example, gravitational particle production \cite{Graham:2015rva, Ahmed:2019mjo, Ema:2019yrd, Ahmed:2020fhc, Kolb:2020fwh}, freeze-in mechanism \cite{Duch:2017khv, Barman:2020ifq, Barman:2021qds}, energy transfer from scalar fields \cite{Agrawal:2017eqm, Agrawal:2018vin, Dror:2018pdh, Co:2018lka, Bastero-Gil:2018uel, Salehian:2020asa} and misalignment mechanism \cite{Nelson:2011sf, Arias:2012az, Nakayama:2019rhg, Kitajima:2023fun}.\footnote{The misalignment mechanism proposed in \cite{Nelson:2011sf} can not produce sufficient DM, which can be fixed by adding a nonminimal coupling to gravity \cite{Arias:2012az}. However, the added nonminimal coupling induces a ghost instability and a quadratic divergence \cite{Himmetoglu:2009qi, Himmetoglu:2008zp, Himmetoglu:2008hx, Esposito-Farese:2009wbc, Graham:2015rva}. A viable misalignment mechanism for VDM can be constructed by including a nonminimal kinetic coupling to the inflaton \cite{Nakayama:2019rhg, Kitajima:2023fun}.}

The wavelike nature of ultralight DM is useful for addressing the cusp-core problem and has interesting phenomenological consequences in cosmology (see \cite{Hu:2000ke, Hui:2016ltb, Ferreira:2020fam} for reviews). Due to their long de Broglie wavelengths, these ultralight bosons suppress structure formation at small scales and affect Jeans instability, baryon acoustic oscillations, halo formation, virialization and interference pattern in galaxies, etc. Compared to their scalar counterparts, ultralight vector bosons have multiple intrinsic degrees of freedom, and their polarizations around the Earth can potentially be probed by terrestrial experiments \cite{Caputo:2021eaa}. On another note, the increased degrees of freedom in VDM reduce the amount of interference and lead to fewer extreme (underdense and overdense) regions \cite{Amin:2022pzv}. As a result, some constraints on VDM mass are slightly weaker than those for scalar DM \cite{Amin:2022pzv}.

While most of the studies on VDM consider vector fields that are minimally coupled to gravity, additional nonminimal couplings are well motivated as they naturally arise as quantum corrections to a minimally coupled classical theory \cite{Birrell:1982ix}. It has been demonstrated that the inclusion of the nonminimal coupling is integral to the renormalization of field theories in curved spacetime \cite{Weinberg:1995mt, callan1970new, freedman1974energy, FREEDMAN1974354}. Moreover, the nonminimal terms are phenomenologically relevant and have been employed in the context of inflation \cite{Turner:1987bw, Ford:1989me, Faraoni:2000wk, Golovnev:2008cf, Golovnev:2008hv, Golovnev:2009ks, Golovnev:2009rm}, modified gravity \cite{Moffat:2005si, Brownstein:2005zz, Tasinato:2014eka, Heisenberg:2014rta, DeFelice:2016yws, deFelice:2017paw}, dark matter \cite{Ji:2021rrn, Sankharva:2021spi, Barman:2021qds, Ivanov:2019iec} and dark energy \cite{Kouwn:2015cdw, Koivisto:2008xf}. 

In this paper, we study wavelike VDM that is nonminimally coupled to gravity in the wave regime, where the mass of DM particles is $m\lesssim 30\mathrm{eV}$. Although the DM mass $m\sim 10^{-22}\mathrm{eV}$ is preferred in some galaxies \cite{Schive:2014dra, Chen:2016unw, Khelashvili:2022ffq}, a lower bound $m \gtrsim 10^{-19} \mathrm{eV}$ has been reported by constraining velocity dispersion in Segue 1 and Segue 2 galaxies \cite{Dalal:2022rmp}, and it further increases to $m\gtrsim 10^{-18}\mathrm{eV}$ if the DM is produced after inflation via a process with a finite-correlation length \cite{Amin:2022nlh}. The impact of nonminimal couplings on these constraints warrants a detailed analysis, and for generality we focus on $m\gtrsim 10^{-22}\mathrm{eV}$. Without investigating the ultraviolet physics, we derive a nonrelativistic effective field theory (EFT) and the modified version of the Schroedinger-Poisson-Friedmann (SPF) equations that include the nonminimal couplings. With this tool, we analyze the impact of nonminimal couplings on vector solitons and the growth of linear perturbations. Beyond scalar perturbations, we show that the nonminimal coupling changes the speed of gravitational waves (GWs), which is used to constrain the parameter space.

The rest of the paper is organized as follows. In section \ref{sec:theory}, we derive the nonrelativistic EFT for nonminimally coupled VDM. In section \ref{sec:phenomenology}, we discuss phenomenological implications of the nonminimal coupling in the context of vector solitons, growth of linear perturbations and GWs. Finally, we summarize our results and suggest possible future directions in section \ref{sec:conclusion}. Throughout the paper, we use the natural units in high-energy physics, $c=\hbar=1$. The reduced Planck mass is defined as $\MP \equiv (8\pi G)^{-1/2}$. Repeated indices are summed unless otherwise stated.

\section{Theoretical setup}
\label{sec:theory}

The nonrelativistic dynamics of wave DM can be described by the SPF equations \cite{Hu:2000ke}, which are the leading-order approximation of the Klein-Gordon-Einstein system. To investigate the impacts of nonminimal couplings to gravity, we will derive a modified version of the SPF equations by identifying small, dimensionless quantities and expanding the relativistic equations to the leading order \cite{Salehian:2020bon, Salehian:2021khb}. This approach can be also extended to allow for a systematic calculation of relativistic corrections \cite{Salehian:2021khb}. Next, we will convert the modified SPF equations to a fluid description, which is more commonly used in the study of linear perturbations.

\subsection{Wave description}
\label{sec:nonrelativistc_eft}

For the purpose of describing the low-energy limit of VDM, we can start without loss of generality from the covariant action $S$ that contains nonminimal couplings to gravity with the lowest mass dimension. The progenitor action we consider is 
\begin{align}
    \label{eq:full_action}
    S = S_G + S_M ~,
\end{align}
where the gravity and matter parts are given by
\begin{align}
	S_G &= \int \dd[4]{x} \sqrt{-g} \left[ \frac{1}{2} \MP^2 R \right] ~, \\
	S_M &= \int \dd[4]{x} \sqrt{-g} \left[ -\frac{1}{4} X_{\mu\nu} X^{\mu\nu} - \frac{1}{2}m^2 X_\mu X^\mu + \frac{1}{2} \xi_1 R X_\mu X^\mu + \frac{1}{2} \xi_2 R^{\mu\nu} X_\mu X_\nu + \cdots \right] ~.
\end{align}
Here, $\xi_1$ and $\xi_2$ characterize the nonminimal coupling to gravity, $R$ and $R_{\mu\nu}$ are the Ricci scalar and Ricci tensor, $X_\mu$ is the VDM field, and $X_{\mu\nu} \equiv \partial_\mu X_\nu - \partial_\nu X_\mu$. In the low-energy limit, an expanding universe containing VDM can be described by the perturbed Robertson-Walker metric \cite{Cembranos:2016ugq}
\begin{align}
	\label{metric}
	\dd{s}^2 = -(1+2\Phi) \dd{t}^2 + a^2(t) (1-2\Phi) \delta\indices{_i_j} \dd{x}^i \dd{x}^j ~.
\end{align}
The action \eqref{eq:full_action} and metric \eqref{metric} are valid if the VDM has become the dominant and nonrelativistic component of the universe.  After the universe enters the matter-dominated era, the dynamics of $X_\mu$ is dominated by oscillations of frequency $\omega\sim m$, thus it is motivated to redefine the vector field in terms of a new, complex nonrelativistic field $\psi_\mu$ by
\begin{align}
	\label{eq:nr_expansion}
	X_\mu(t,\vb{x}) = \frac{1}{\sqrt{2ma}} \left[ e^{-imt} \psi_\mu(t,\vb{x}) + e^{imt} \psi_\mu^*(t,\vb{x}) \right] ~.
\end{align}
The power of the scale factor is chosen such that the amplitude of $\psi_\mu$ does not change significantly with the expansion of the universe, since the energy density scales like $\rho\sim m^2 a^{-2} X_i X_i \propto a^{-3}$ during matter domination. To ensure that the field redefinition preserves the number of propagating degrees of freedom, one could employ a constraint that implies a field equation of $\psi_\mu$ remaining first order in time derivatives \cite{Salehian:2020bon},
\begin{align}
	\label{eq:nr_constraint}
	e^{-imt} \dot\psi_i + e^{imt} \dot\psi_i^* = 0 ~.
\end{align}
An alternative nonlocal field redefinition was exploited in \cite{Namjoo:2017nia}.

To derive a nonrelativistic EFT of VDM, we follow the prescription in \cite{Salehian:2020bon, Salehian:2021khb} and identify the following small, dimensionless parameters
\begin{align}
	\epsilon_H \sim \frac{H}{m} ~,\quad
	\epsilon_t \sim \abs{ \frac{\dot Q}{m Q} } ~,\quad
	\epsilon_k \sim \abs{ \frac{\nabla^2 Q}{a^2 m^2 Q} } ~,\quad
	\epsilon_\psi \sim \frac{|\psi_i|}{\MP \sqrt{m}} ~,\quad
	\epsilon_g \sim |\Phi| ~,
\end{align}
where $Q$ can be any of the slowly varying variables including $a, H, \psi_\mu, \Phi$. These parameters respectively characterize the smallness of the expansion rate, time variation, spatial gradient, field amplitude and gravity strength. In addition, the nonminimal coupling terms in \eqref{eq:full_action} should not play too significant a role in order to retain the success of General Relativity and to avoid the ghost instability of longitudinal modes discussed in refs.~\cite{Himmetoglu:2009qi, Himmetoglu:2008zp, Himmetoglu:2008hx, Esposito-Farese:2009wbc, Graham:2015rva}. Hence one must have another small parameter much less than unity in terms of $\xi_a ~(a=1,2)$,
\begin{align}
	\epsilon_\xi \sim \abs{ \frac{\xi_a R}{m^2} } ~,
\end{align}
where $R\sim a^{-2} \nabla^2\Phi \sim \rho/\MP^2$ and $\rho$ is the local DM density. This sets an upper limit on $\xi_a$,
\begin{align}
	\label{xi_range}
	\abs{\xi_a} \ll \frac{m^2 \MP^2}{\rho} = 1.5\times 10^{15} \left( \frac{m}{10^{-20} \mathrm{eV}} \right)^2 \left( \frac{5 \times 10^{10} ~\mathrm{GeV/m^3}}{\rho} \right) ~.
\end{align}
Here the reference value of DM density is taken to be its average value at the matter-radiation equality \cite{ParticleDataGroup:2022pth}. If one is interested in the dynamics of local DM today, one should take the DM density to be around $0.3~\mathrm{GeV/cm^3}$ \cite{deBoer:2010eh, Bovy:2012tw, mckee2015stars, Sivertsson:2017rkp}, which implies $\abs{\xi_a} \ll 2.6\times 10^{20} (m / 10^{-20} \mathrm{eV})^2$.\footnote{In section \ref{sec:gw_speed}, we will take this value as the upper limit of $\xi_a$ when studying the variation of gravitational wave speed.} In general, these small parameters are not independent from each other and we do not know a priori the relative magnitudes between them, thus a reliable expansion strategy would require the system of equations to be expanded up to a specific order that includes every parameter. That is, each order is a homogeneous function of all small parameters. In the following discussions, we will denote all small parameters collectively by $\epsilon = \{\epsilon_H, \epsilon_t, \epsilon_k, \epsilon_\psi, \epsilon_g, \epsilon_\xi \}$ for notation convenience and work up to the leading order in $\epsilon$.

By plugging the field redefinition \eqref{eq:nr_expansion} into the action \eqref{eq:full_action} and integrating out the fast oscillating terms, we obtain a nonrelativistic effective action
\begin{align}
	\label{eq:nr_action}
	S = \int \dd[4]{x} & \bigg[ \MP^2 a\left( -3 \dot a^2 + \Phi\nabla^2\Phi - 6a \ddot a\Phi \right) \nonumber \\
	& + \frac{1}{2} a^2 m \abs{\psi_0 - \frac{i}{a^2m}\nabla\cdot\vb*{\psi}}^2 + i \dot{\vb*{\psi}}\cdot\vb*{\psi}^\ast + \frac{1}{2a^2m} (\nabla^2\vb*{\psi})\cdot\vb*{\psi}^* - m\Phi \abs{\vb*{\psi}}^2  \nonumber \\
	& + \frac{\abs{\vb*{\psi}}^2}{2a^2m} \left[ 2 \xi_1 (\nabla^2 \Phi + 3 \dot{a}^2 + 3a\ddot a ) + \xi_2 (\nabla^2\Phi + 2 \dot{a}^2 + a\ddot a) \right]\bigg] \times \left[ 1 + \order{\epsilon} \right] ~,
\end{align}
where $\vb*{\psi}$ is the spatial part of $\psi_\mu$, the overdot stands for time derivative, and $\order{\epsilon}$ includes all relativistic corrections.\footnote{Higher-order time derivatives will emerge if one takes the relativistic corrections into account, but they do not introduce unphysical degrees of freedom because they can be systematically removed by applying field equations at lower orders \cite{Namjoo:2017nia}.} To the leading order, the constraint equation on $\psi_0$ is not affected by the nonminimal coupling, implying that the existence of nonminimal couplings does not bring about the singularity problem discussed in \cite{Mou:2022hqb, Clough:2022ygm, Coates:2022qia}. As promised, \eqref{eq:nr_action} is also free from the ghost instability of longitudinal modes \cite{Himmetoglu:2009qi, Himmetoglu:2008zp, Himmetoglu:2008hx, Esposito-Farese:2009wbc, Graham:2015rva}.

By varying \eqref{eq:nr_action} with respect to $a,\psi_i^*,\Phi$, in the nonrelativistic limit, we obtain the field equations
\begin{align}
	\label{EOM_psi}
	i\partial_t \psi_i &= - \frac{\nabla^2}{2m a^2} \psi_i + m\Phi_N \psi_i + 2m\Phi_\xi \psi_i - \frac{(2\xi_1+\xi_2)\nabla^2\Phi_\xi}{2m a^2} \psi_i ~,\\
	\label{EOM_Phi}
	\frac{\nabla^2}{a^2}\Phi_N &= 4\pi G (\rho-\bar\rho) ~,\quad
	\Phi_\xi = -\frac{(2\xi_1+\xi_2)2\pi G}{m^2}\rho ~,\\
	\label{EOM_H}
	H^2 &= \frac{8\pi G}{3} \bar\rho ~,\quad
	\rho = \frac{1}{a^3} m \abs{\boldsymbol\psi}^2 \equiv \sum_{i=1}^{3} \rho_i ~,
\end{align}
where $\Phi = \Phi_N + \Phi_\xi$, the overline stands for spatial averaging and we have broken the energy density into its component parts $\rho_i \equiv a^{-3} m \abs{\psi_i}^2$. Besides the Newtonian potential $\Phi_N$, the nonminimal coupling results in a nonminimal potential $\Phi_\xi$, which represents attractive (repulsive) self-interactions for $2\xi_1+\xi_2>0$ ($<0$). These self-interactions do not depend on specific polarization states, and are distinct from those due to terms of the form $(X_\mu X^\mu)^2$, which would result in both $\psi_j \psi_j^\ast \psi_i$ and $\psi_j \psi_j \psi_i^\ast$ terms in \eqref{EOM_psi} and can break the degeneracy in energy of polarized vector solitons \cite{Jain:2021pnk, Zhang:2021xxa}. The modified SPF equations \eqref{EOM_psi}-\eqref{EOM_H} are our master equations in the wave description for nonrelativistic VDM.

\subsection{Fluid description}
The fluid description is related to the SPF equations through the Madelung transformation \cite{madelung1927quantentheorie}. To do the transform, let us define the fluid velocity ${\boldsymbol v}_i$ in terms of the phase of each field component $\psi_i$,
\begin{align}
	\psi_i \equiv \sqrt{\frac{\rho_i a^3}{m}} e^{i\theta} ~,\quad
	{\boldsymbol v}_i \equiv \frac{1}{ma} \nabla\theta_i ~.
\end{align}
For convenience, we do not use the Einstein summation convention in and only in this subsection. The equations \eqref{EOM_psi} become
\begin{align}
	\label{EOM_continuity}
	\dot\rho_i + 3 H \rho_i +  \frac{1}{a} \nabla\cdot(\rho_i {\boldsymbol v}_i) &= 0 ~,\\
	\label{EOM_Euler}
	\dot{{\boldsymbol v}_i} + H{\boldsymbol v}_i + \frac{1}{a} ({\boldsymbol v}_i\cdot\nabla) {\boldsymbol v}_i &= - \frac{1}{a}\nabla \left(\Phi_N + \Phi_{Q,i} + 2\Phi_\xi - \frac{(2\xi_1+\xi_2)}{2m^2a^2}\nabla^2\Phi_\xi \right) ~,
\end{align}
where $\Phi_\xi$ is given by \eqref{EOM_Phi} and
\begin{align}
	\label{quantum_potential}
	\Phi_{Q,i} \equiv - \frac{1}{2a^2m^2} \frac{\nabla^2\sqrt{\rho_i}}{\sqrt{\rho_i} } ~.
\end{align}
Physically, the Newtonian potential $\Phi_N$ attracts matter, the quantum potential $\Phi_Q$ repulses matter, and the property of interactions induced by nonminimal couplings depends on the sign of $2\xi_1+\xi_2$. In the large mass limit, the equations can be used to describe particle-like cold dark matter (CDM) \cite{Widrow:1993qq}. In section \ref{sec:perturbation}, we will use \eqref{EOM_continuity} and \eqref{EOM_Euler} to study the growth of linear perturbations and the Jeans scale.

\section{Phenomenological implications}
\label{sec:phenomenology}

With the wave and fluid description of VDM, we will study the mass-radius relation of vector solitons and the growth of linear perturbations in section \ref{sec:solitons} and \ref{sec:perturbation} respectively. Moving beyond scalar perturbations, we will investigate the effects of nonminimal couplings on GW speed in section \ref{sec:gw_speed}.

\subsection{Mass-radius relation of vector solitons}
\label{sec:solitons}

Solitons are stable (or long-lived) and spatially localized field configurations that have been extensively studied for scalar fields over the past several decades \cite{kaup1968klein, ruffini1969systems, Friedberg:1976me, Coleman:1985ki, Lee:1988av, Seidel:1991zh, Copeland:1995fq, Alcubierre:2003sx, Amin:2010jq, Hertzberg:2010yz, Amin:2011hj, Grandclement:2011wz, Salmi:2012ta, Schive:2014dra, Lozanov:2016hid, Hui:2016ltb, Visinelli:2017ooc, Helfer:2018vtq, Levkov:2018kau, Amin:2019ums, Olle:2019kbo, Sanchis-Gual:2019ljs, Niemeyer:2019gab, Zhang:2020bec, Zhang:2020ntm, Dmitriev:2021utv, Croft:2022bxq}; see refs.~\cite{Lee:1991ax, Liebling:2012fv, Nugaev:2019vru} for reviews.  Unlike their scalar counterparts, vector solitons have started to attract attention only more recently, and they have been shown to admit three types of classically stable solutions with zero orbital angular momentum \cite{Loginov:2015rya, Brito:2015pxa, Sanchis-Gual:2017bhw, Adshead:2021kvl, Jain:2021pnk, Zhang:2021xxa, Gorghetto:2022sue, Amin:2022pzv, Amin:2023imi, Jain:2023ojg}. Here we will calculate the mass-radius relation of the ground state solitons, which are not spherically symmetric in field configuration but are spherically symmetric in energy/mass density \cite{Jain:2021pnk, Zhang:2021xxa}. As the size of these solitons is much smaller than the Hubble patch, we will neglect the expansion of the universe and set $a(t)=1$.

The mass-radius relation can be found by minimizing the energy of the soliton at a fixed particle number. We define the total particle number $N$ and mass $M$ of a vector soliton by
\begin{align}
	\label{soliton_number_mass}
	N \equiv \int \dd[3]{x} \psi_i \psi_i^* ~,\quad
	M \equiv \int \dd[3]{x} \rho = m N ~,
\end{align}
where $N$ is the conserved charge associated with the global $U(1)$ symmetry $\psi_i\rightarrow e^{i\alpha} \psi_i$ in the nonrelativistic action \eqref{eq:nr_action}. The radius $R$ is defined as that of the ball enclosing $99\%$ of the total mass. The nonrelativistic action \eqref{eq:nr_action} is also invariant under time translation, and the associated conserved charge is identified as the energy $E_\psi$,
\begin{align}
	\label{eq:soliton_energy}
	E_\psi = \int \dd[3]{x} \left[ \frac{1}{2m} \partial_j \psi_i \partial_j \psi_i^* + m\Phi_N \abs{\boldsymbol\psi}^2 + 2m\Phi_\xi \abs{\boldsymbol\psi}^2 - \frac{(2\xi_1 + \xi_2) \nabla^2\Phi_\xi}{2m} \abs{\boldsymbol\psi}^2 \right] ~.
\end{align}
This energy does not include the rest mass energy of the original vector field $X_\mu$. In the nonrelativistic limit, the total energy of $X_\mu$ is $E_X=E_\psi + mN$. Using the thin-wall approximation and replacing $\nabla$ with $1/R$ and $\int \dd[3]{x}$ with $R^3$ \cite{Lee:1991ax, Coleman:1985ki, Visinelli:2017ooc}, the energy can be written as\footnote{An alternative method is to assume a Gaussian ansatz for soliton profiles \cite{Chavanis:2011zm}. However, numerically Gaussian profiles are bad approximations if the nonminimal coupling becomes important.}
\begin{align}
	\label{eq:E_M_R_relation}
	E_\psi \propto \frac{\beta_1}{2} \frac{M}{m^2 R^2} - \frac{M^2}{\MP^2 R} - \frac{\beta_2}{3} \frac{(2\xi_1+\xi_2) M^2}{m^2 \MP^2 R^3} - \frac{\beta_3}{5} \frac{(2\xi_1+\xi_2)^2 M^2}{m^4 \MP^2 R^5} ~,
\end{align}
where $\beta_1,\beta_2,\beta_3$ are numerical coefficients to be determined, and we have replaced center field amplitudes $\psi_c$ with $M\sim m\psi_c^2 R^3$. This expression of energy allows us to minimize the energy at a fixed particle number, to wit, $\delta(E_\psi+\mu N)=0$ where $\mu$ is a Lagrangian multiplier \cite{Lee:1991ax}. By varying $E_\psi+\mu N$ with respect to $R$ for a fixed $M$, we obtain
\begin{align}
	\label{eq:MR}
	M = \frac{\beta_1 m^3R^3}{m^4R^4 + \beta_2(2\xi_1+\xi_2) m^2R^2 + \beta_3 (2\xi_1+\xi_2)^2} \frac{\MP^2}{m} ~.
\end{align}
The numerical coefficients $\beta_1,\beta_2,\beta_3$ can be determined by comparing \eqref{eq:MR} with the mass-radius relation obtained using the numerical shooting method, which is outlined in appendix \ref{app:soliton_shooting}. 
\begin{figure}
	\centering
	\includegraphics[width=0.6\linewidth]{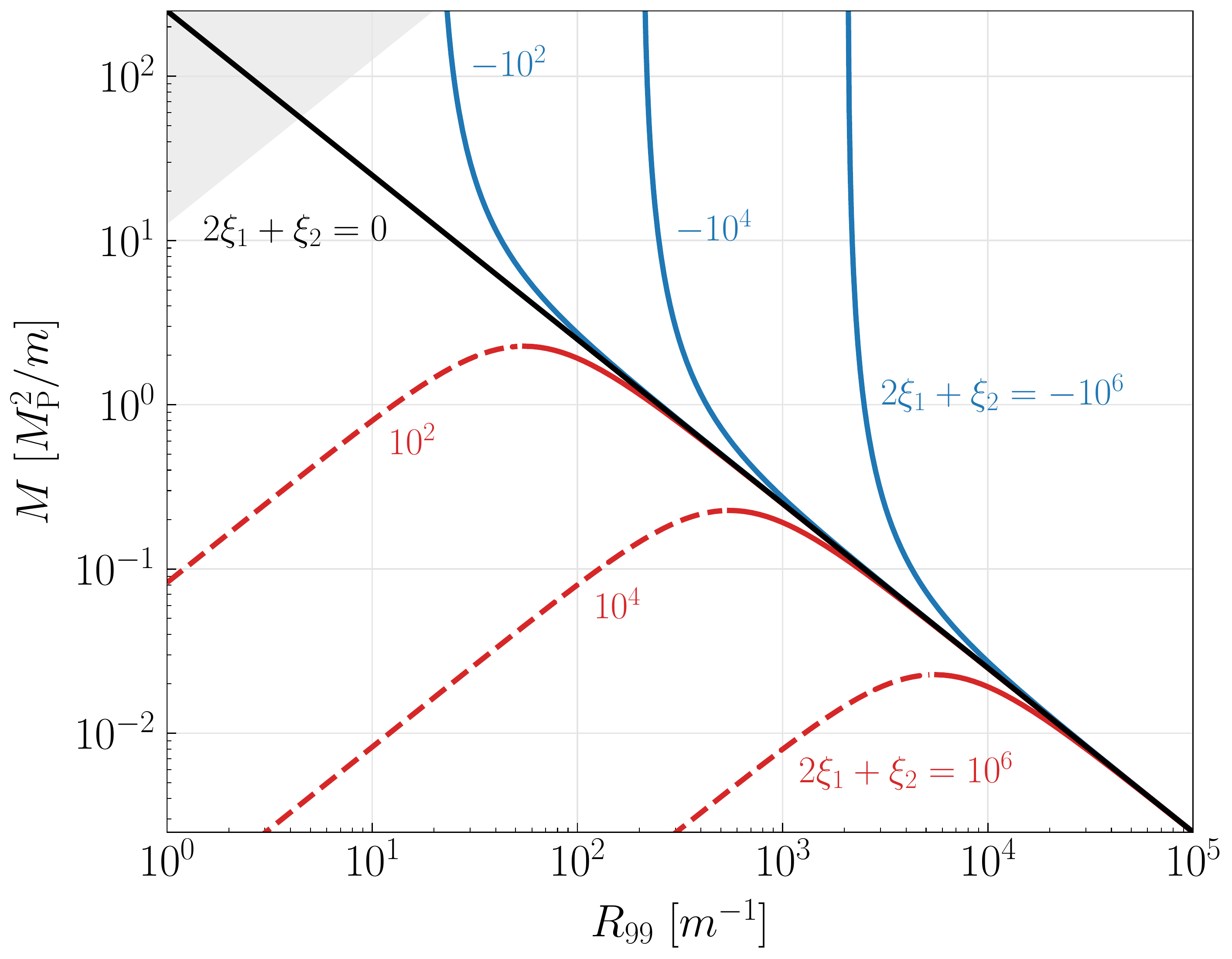}
	\caption{Mass-radius relation of the ground-state vector solitons, given by equation \eqref{eq:MR}. The dashed line indicates a classical instability against small perturbations. The gray region at the upper left corner corresponds to the regime in which $R$ is comparable with the Schwarzschild radius, $mM/\MP^2 \leq 4\pi mR$, where gravity becomes strong and the nonrelativistic approximation breaks down.}
	\label{fig:mass_radius_relation}
\end{figure}
We find the above formula to be a good approximation with $\beta_1=250$, $\beta_2=30.25$, $\beta_3=0$ for $2\xi_1+\xi_2>0$ and with $\beta_1=250$, $\beta_2=8.5$, $\beta_3=18.06$ for $2\xi_1+\xi_2<0$. The mass-radius relation \eqref{eq:MR} is shown in figure \ref{fig:mass_radius_relation}, which manifests a similar qualitative behavior compared to that from covariant quartic self-interactions \cite{Chavanis:2011zi, Chavanis:2011zm, Salehian:2021khb}. One can see that the low-radius part of the curve is dramatically changed compared to the minimal case, signifying the fact that the nonminimal coupling becomes more important than Newtonian gravity at small scales.

We end this subsection with a brief discussion on the possibility of inferring soliton configurations from the density profile of galaxies. Solitons offer a natural solution to the core-cusp problem, which refers to the mismatch between the cuspidal density profile predicted by CDM simulations and the flatter ones observed in galactic centers \cite{Ferreira:2020fam}. By hosting a solitonic core, the galaxy can exhibit a smooth central density profile, thereby preventing gravitational clustering of matter and resolving the core-cusp problem. While a proliferation of favorable evidence for such solitonic cores has emerged in recent years, the existence of solitonic cores also pose strong constraints on the mass of dark matter bosons, which were obtained by comparing the predicted density profiles of solitonic galaxy cores and the observed ones \cite{Schive:2014hza, Marsh:2015wka, Chen:2016unw, Gonzalez-Morales:2016yaf, Broadhurst:2019fsl, Bar:2018acw, Veltmaat:2018dfz, DeMartino:2018zkx, Pozo:2023zmx, ParticleDataGroup:2022pth}. Nonminimal couplings to gravity could weaken such constraints and alleviate tensions between the halo mass and soliton mass in some galaxies \cite{Robles:2018fur, Desjacques:2019zhf, Safarzadeh:2019sre}.  Further investigation is needed to explore this possibility, and we leave that for future work.

\subsection{Growth of linear perturbations}
\label{sec:perturbation}

In this section, we will use the fluid description to study the growth of density perturbations for wave VDM.\footnote{For scalar DM, the density perturbation has been calculated in the fluid description up to the third order in $\delta$ and $v$ to obtain the one-loop power spectrum \cite{Li:2018kyk}. One may also use the wave perturbations $\delta\psi\equiv \psi-\bar\psi$ to study structure growth; that said, this approach breaks down at higher redshifts compared to the fluid description \cite{Li:2018kyk}.} We find that significant nonminimal couplings may lead to distinct evolution of perturbations, where both the small and large scale perturbations can grow significantly. As the universe expands, the importance of the nonminimal coupling decreases and eventually the standard picture of wave DM \cite{Hu:2000ke} is recovered, albeit with a modified Jeans scale.

Linearizing the fluid equations \eqref{EOM_continuity} and \eqref{EOM_Euler}, we obtain the equations for perturbations around a homogeneous background
\begin{align}
	\dot\delta_i + \frac{1}{a}\nabla\cdot{\boldsymbol v}_i &= 0 ~,\\
	\dot{{\boldsymbol v}_i} + H{\boldsymbol v}_i  &= -\frac{1}{a}\nabla \left(\Phi_N + \Phi_{Q,i} + 2 \Phi_\xi - \frac{(2\xi_1+\xi_2)}{2m^2a^2}\nabla^2\Phi_\xi \right) ~,
\end{align}
where $\Phi_N,\Phi_{Q,i},\Phi_\xi$ are given by \eqref{EOM_Phi} and \eqref{quantum_potential}, $\rho_i \equiv \bar\rho_i (1+\delta_i)$ and the background densities satisfy $\dot{\bar\rho}_i + 3H\bar\rho_i = 0$. These equations can be combined to yield a set of second-order differential equations for the overdensity $\delta_i$. In Fourier space, it can be written as
\begin{align}
	\label{overdensity_eq}
	\left[\partial_t^2 + 2H\partial_t +
	\begin{pmatrix}
		\frac{k^4}{4a^4m^2} + c_1 & c_2 & c_3 \\
		0 & \frac{k^4}{4a^4m^2} & 0 \\
		0 & 0 & \frac{k^4}{4a^4m^2}
	\end{pmatrix}
	\right]
	\begin{pmatrix}
		\Delta_1 \\
		\Delta_2 \\
		\Delta_3
	\end{pmatrix} = 0 ~,
\end{align}
where
\begin{align*}
	\begin{pmatrix}
		\Delta_1 \\
		\Delta_2 \\
		\Delta_3
	\end{pmatrix}
	\equiv
	Q
	\begin{pmatrix}
		\delta_1 \\
		\delta_2 \\
		\delta_3
	\end{pmatrix} ~,\quad
	Q = \frac{1}{\sqrt{6}} 
	\begin{pmatrix}
		\sqrt{2} & \sqrt{2} & \sqrt{2} \\
		2 & -1 & -1 \\
		0 & \sqrt{3} & -\sqrt{3}
	\end{pmatrix} ~,
\end{align*}
and we have defined
\begin{align*}
	c_1 &\equiv - \left[ \frac{k^2}{2m^2a^2} (2\xi_1+\xi_2) \sqrt{4\pi G\bar\rho} + \sqrt{4\pi G\bar\rho} \right]^2 ~,\\
	c_2 &\equiv -\frac{\pi G (2a^2m^2 + k^2(2\xi_1+\xi_2))^2}{\sqrt{2} a^4m^4} (2\bar\rho_1 - \bar\rho_2 - \bar\rho_3) ~,\\
	c_3 &\equiv -\frac{\sqrt{3}\pi G (2a^2m^2 + k^2(2\xi_1+\xi_2))^2}{\sqrt{2} a^4m^4} (\bar\rho_2 - \bar\rho_3) ~.
\end{align*}
Assuming $\bar\rho_1=\bar\rho_2=\bar\rho_3$, which is expected unless there is a mechanism that favors VDM with a particular polarization state at the time of production and throughout its subsequent evolution, the differential operator of \eqref{overdensity_eq} becomes diagonal and one can study $\Delta_i$ separately.\footnote{Since $\Delta_1$ corresponds to the direction $\delta_1 = \delta_2 = \delta_3$ (which does not distinguish between spatial indices), $\Delta_1$ should be interpreted as the isotropic component of the density contrast; accordingly, $\Delta_2$ and $\Delta_3$ are the anisotropic components. If $\bar{\rho}_1 = \bar{\rho}_2 = \bar{\rho}_3$, then equation \eqref{overdensity_eq} is symmetric under permutations of the spatial indices $i=1,2,3$, and the isotropic and anisotropic components correspond to orthogonal invariant subspaces of the solution space. Note that the orthogonal transform $Q$ is not to be understood as a spatial rotation since the $\delta_i$'s do not form a vector under spatial rotation.} While both $\Delta_2$ and $\Delta_3$ oscillate and do not experience significant growth due to their positive effective mass, the growth of $\Delta_1$ depends on the sign of 
\begin{align}
	\label{jeans_effective_mass}
    \nonumber
	\Omega^2 &\equiv \frac{k^4}{4a^4m^2} + c_1 \\
    \nonumber
    &= -4\pi G \bar\rho \left[ 1 + (2\xi_1 + \xi_2) \frac{k^2}{a^2m^2}  + \left( (2\xi_1 + \xi_2)^2 - \frac{m^2}{4\pi G \bar\rho} \right) \frac{k^4}{4a^4m^4} \right] \\
    &= \frac{1}{4a^4m^2} \left[ 1 -(2\xi_1+\xi_2)^2 \frac{4\pi G \bar{\rho}}{m^2} \right] (k^2 + k_-^2)(k^2 - k_+^2) ~,
\end{align}
where
\begin{align}
	k_\pm^2 \equiv k_{J,0}^2 \frac{1}{1 \mp (2\xi_1 + \xi_2)\sqrt{4\pi G\bar\rho/m^2}} ~,\quad
	k_{J,0} = (16\pi G m^2 \bar\rho a^4)^{1/4} ~,
\end{align}
with $k_{J,0}$ being the comoving Jeans scale without nonminimal couplings \cite{Hu:2000ke, Gorghetto:2022sue}. The sign of $\Omega^2$ is undetermined, because the upper limit of $\xi_a$ $(a=1,2)$ is given by $\xi_a\ll m^2 \MP^2 / \bar\rho$, i.e. equation \eqref{xi_range}, which does not fix the relative magnitude between $\xi_a^2$ and $m^2 \MP^2 / \bar\rho$.

The nonminimal couplings can play an important role if $(2\xi_1 + \xi_2)^2\gg m^2\MP^2/\bar\rho$. In this case, we have $k_+^2<0, k_-^2>0$ for positive $2\xi_1+\xi_2$, hence $\Omega^2<0$ and perturbations grow for all $k$ modes.  For negative $2\xi_1+\xi_2$, we have $k_+^2>0, k_-^2<0$, thus perturbations with $\abs{k_+}<k<\abs{k_-}$ oscillate whereas others grow. In contrast, the small-scale perturbations with $k>k_{J,0}$ are suppressed for minimally coupled wave DM \cite{Hu:2000ke, Gorghetto:2022sue} -- the presence of significant nonminimal couplings could enhance small-scale structure.

The evolution of CDM perturbations $\delta \propto a$ for $k<k_{\rm obs} \sim 10 ~h\mathrm{Mpc^{-1}}\sim 10^3 k_{\rm eq}$ is consistent with current observations \cite{Irsic:2017yje, Bechtol:2022koa}. To retain the success of CDM on large-scale perturbations, we demand $\Omega^2 \approx -4\pi G\bar\rho$ for $k<k_{\rm obs}$ since the matter-radiation equality, which yields
\begin{align}
	\label{xi_large_scale_perturbation}
	\abs{2\xi_1 + \xi_2} \ll \frac{a_{\rm eq}^2m^2}{k_{\rm obs}^2} = 10^{10} \left( \frac{m}{10^{-20} \mathrm{eV}} \right)^2 \left( \frac{10^{-28} \rm{eV}}{H_\mathrm{eq}} \right)^2 ~,
\end{align}
This bound is more stringent than equation \eqref{xi_range}, but weaker than the constraint we will obtain for $\xi_2$ in section \ref{sec:gw_speed} based current limits on GW speed.

As the energy density decreases with the expansion of the universe, eventually $m^2\MP^2/\bar\rho$ will dominate over $(2\xi_1+\xi_2)^2$ and we recover the standard evolution of perturbations for wave DM \cite{Hu:2000ke, Gorghetto:2022sue}. The comoving Jeans scale for the subsequent evolution is
\begin{align}
	\label{jeans_scale}
	k_J = k_+ \simeq a (16 \pi G m^2 \bar\rho)^{1/4} \left[ 1 + (2\xi_1 + \xi_2) \sqrt{\frac{\pi G \bar\rho}{m^2}} \right] ~.
\end{align}
Perturbations grow at large length scales with $k\ll k_J$ and oscillate at small scales with $k\gg k_J$.

It has been shown that minimally coupled wave DM has a sharp break in the power spectrum of density perturbations with $k$ larger than $k_{J,0,\mathrm{eq}}$, the comoving Jeans scale at matter-radiation equality~\cite{Hu:2000ke}. The existence of nonminimal couplings is expected to leave imprints on the spectrum at small scales and also shift the break toward higher or lower $k$ depending on the value of $2\xi_1+\xi_2$. The small-scale structure could be enhanced if the nonminimal coupling is significant in early times. It might be interesting to explore the related phenomenology in detail for future work.

\subsection{Speed of gravitational waves}
\label{sec:gw_speed}
In this subsection, we study the impact of nonminimally coupled VDM on the speed of GWs.
We find that observations regarding GW speed give constraints on parameters $\xi_2$ and $m$, which are summarized in equations \eqref{eq:gw_constraint_LIGO} and \eqref{eq:gw_constraint_Cherenkov}.
Since we only consider GW propagation on scales much smaller than $H_0^{-1}\simeq 4~\mathrm{Gpc}$, we will henceforth ignore the expansion of the universe by setting $a(t)=1$.

A plane GW travelling in the $+z$ direction has two polarization modes $h_+(t,z)$ and $h_\times(t,z)$.
These two modes satisfy a generic wave equation:
\begin{align}
    \partial_t^2 h_\lambda - c_T^2 \partial_z^2 h_\lambda = 0
    ~,
\end{align}
where $c_T$ is the GW speed, and $c_T = 1$ for ordinary linearized gravity.
It is customary to parameterize the correction to GW speed by $\alpha_T \equiv c_T^2 - 1$.
On a coherent nonrelativistic VDM background, we have averaged speed correction:
\begin{align}
    \label{eq:alpha_T}
    \alpha_T = \frac{\xi_2 \rho_3}{m^2 \MP^2}
    ~,
\end{align}
where $\rho_3$ is the mass density contributed by $\psi_3$, the component of $\vb*{\psi}$ in the direction of GW propagation.
See also eq.~\eqref{eq:alpha_T_psi} and appendix \ref{app:GW_speed} for a detailed derivation.
One can see that a non-zero $\xi_2$ can result in a GW speed different from $1$, whereas the $\xi_1$ term does not change the GW speed.\footnote{In the case of scalar wave dark matter~\cite{Ji:2021rrn,Ivanov:2019iec}, nonminimal terms of the form $R \phi^2$ do not change the GW speed, whereas Horndeski type terms of the form $G\indices{_\mu_\nu} \nabla^\mu \phi \nabla^\nu \phi$ do.}

A non-zero $\alpha_T$ can give rise to significant difference between the time it takes for GWs and light to travel over the same path.
If GWs and light are emitted from the same source, then at the end of propagation, the GW signal is delayed from the light signal by $\Delta t$:
\begin{align}
\label{eq:delta_t_general}
    & \Delta{t} 
    = \int \left(\frac{1}{c_T(s)} - 1\right) \dd{s}
    \approx - \frac12 \int \alpha_T(s) \dd{s}
    \approx - \frac{\xi_2}{6 m^2 \MP^2} \int \rho(s) \dd{s}
    ~,
\end{align}
where the integral is over the path of GW/light propagation. 
In the calculation above, we use $|\alpha_T| \sim \epsilon_\xi \ll 1$ for linear approximation, and take $\expval{\rho_3}=\rho/3$ based on the assumption that $\vb*{\psi}$ has randomized polarizations.

For a GW travelling from a faraway galaxy (say, $100~\mathrm{Mpc}$ away) to Earth, the time delay $\Delta t$ is mostly accumulated inside galaxy haloes, including the source galaxy and the Milky Way.
To see this, simply estimate the integral with the respective dark matter density and characteristic length:
\begin{align}
\label{eq:rho_integral_estimate}
\int \rho(s) \dd{s} \approx
\begin{cases}
    (4.6~\mathrm{GeV/m^3}) (100~\mathrm{Mpc}) \approx 6 \times 10^{-7}~\mathrm{GeV}^3 & \text{(Outside galaxies)} \\
    (0.3~\mathrm{GeV/cm^3}) (30~\mathrm{kpc}) \approx 1 \times 10^{-5}~\mathrm{GeV^3} & \text{(Inside Milky Way)} 
\end{cases}
\end{align}
We used the critical density $\rho_c \simeq 4.6~\mathrm{GeV/m^3}$ for the ``Outside galaxies" estimate, which is conservative since the dark matter density outside galaxies should be lower than average.
One can see that, even in this conservative estimate, the time delay $\Delta t$ is primarily accumulated within galaxies.

\begin{figure}[t]
	\centering
	\begin{minipage}{0.49\linewidth}
		\includegraphics[width=\linewidth]{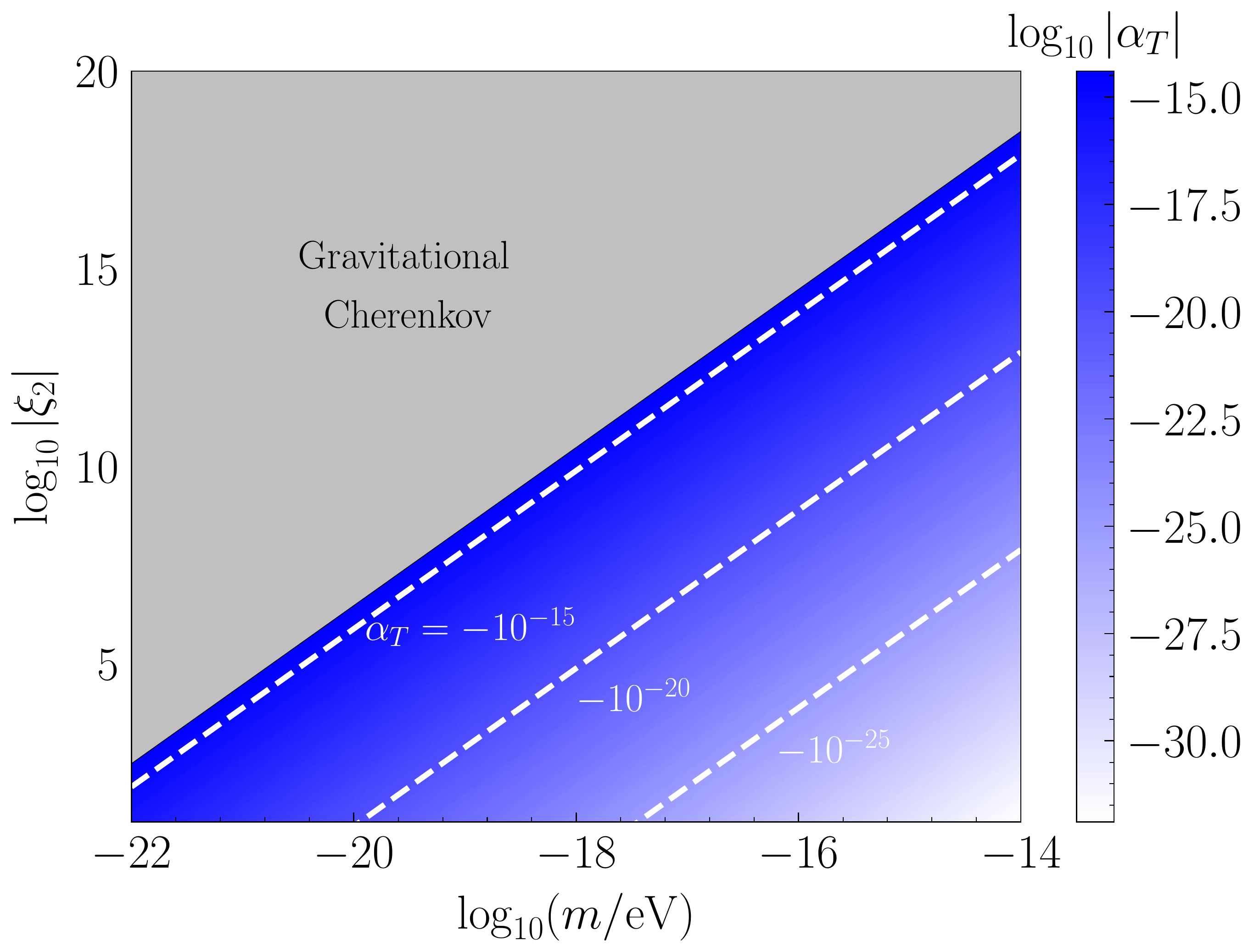}
	\end{minipage}
	\begin{minipage}{0.49\linewidth}
		\includegraphics[width=\linewidth]{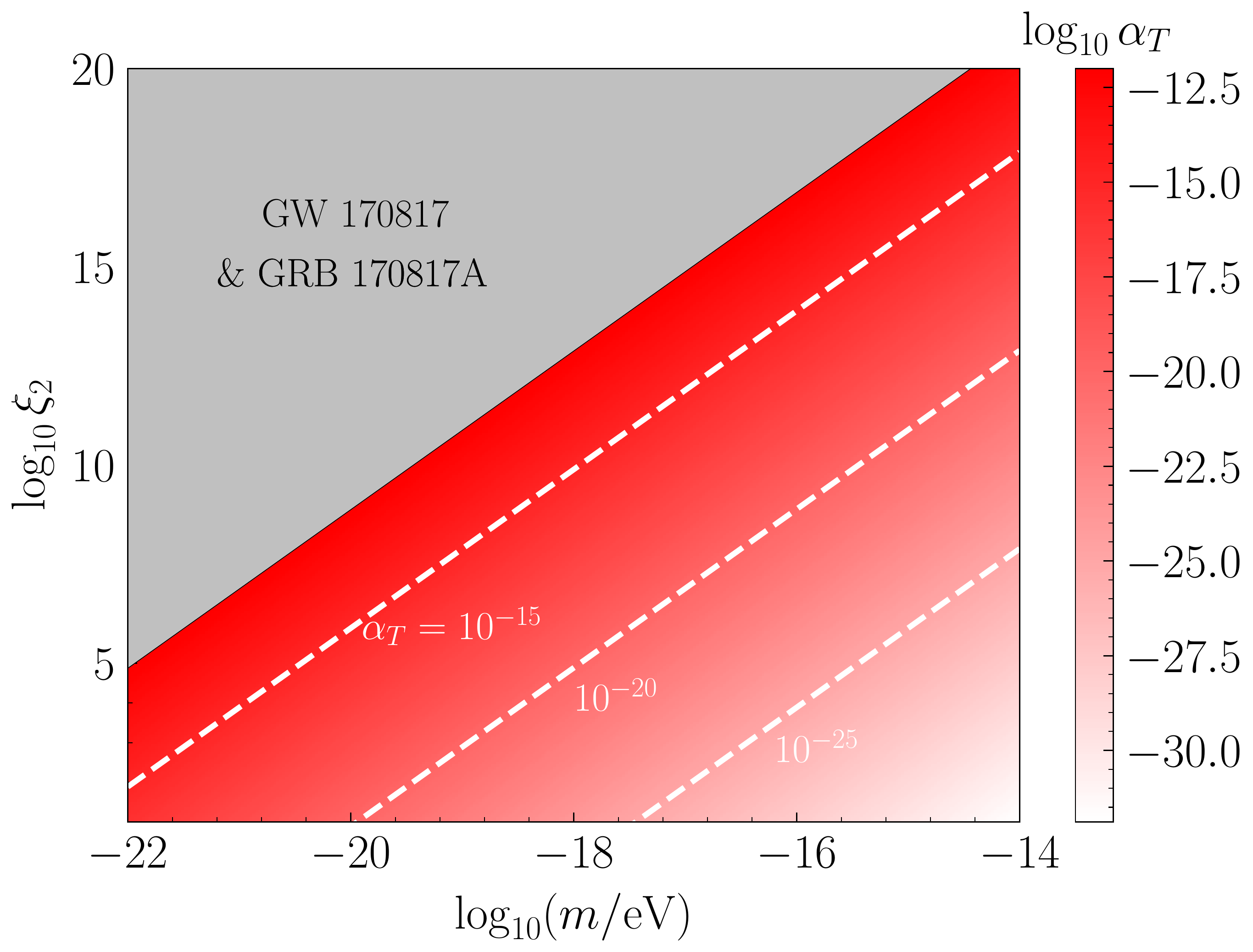}
	\end{minipage}
	\caption{Observational constraints on $m$ and $\xi_2$.  The left panel gives the constraint for $\xi_2 < 0$, and the right panel gives the constraint for $\xi_2 > 0$.  Light gray regions on both panels are excluded.  On the left panel, the constraint comes from the lack of gravitational Cherenkov radiation; on the right panel, the constraint comes from the observation of GW170817 and GRB170817A.  The white dashed lines on both panels are level sets of $\alpha_T$, averaged inside a typical galaxy halo.}
	\label{fig:gw_speed}
\end{figure}

With the formula \eqref{eq:delta_t_general} for time delay, a stringent bound on $\xi_2 / m^2$ is implied by the detection of GW170817 and GRB 170817A, which are the GW and $\gamma$-ray signals produced by a binary neutron star (BNS) merger event~\cite{LIGOScientific:2017zic, Baker:2017hug}.
We denote the time it takes for the BNS system to starting emitting $\gamma$-rays after the merger by $\Delta t_\text{emission} = \order{10~\mathrm{s}} > 0$, the time between receiving the GW signal for merger and detecting the first $\gamma$-rays by $\Delta t_\textrm{detection} = -1.7~\mathrm{s}$, and the time delay due to modified GW speed by $\Delta t$ as in eq.~\eqref{eq:delta_t_general}.
Then $\Delta t = \Delta t_\textrm{detection} + \Delta t_\text{emission}$.
For $\Delta t_\text{emission}$ ranging from $0 \sim 10~\mathrm{s}$, we have $-1.7~\mathrm{s} \leq \Delta t \leq 8.3~\mathrm{s}$, which yields a bound on $\xi_2 / m^2$:
\begin{align}
\label{eq:gw_constraint_LIGO}
    -4 \times 10^{49}~\mathrm{eV^{-2}} 
    \lesssim \frac{\xi_2}{m^2} \lesssim 
    8 \times 10^{48}~\mathrm{eV^{-2}} 
    ~,
\end{align}
where we used the Milky Way estimate in eq.~\eqref{eq:rho_integral_estimate} for the $\rho$ integral.
The constraints are shown in figure \ref{fig:gw_speed}.

A stronger lower bound on $\xi_2 / m^2$ is due to the lack of gravitational Cherenkov radiation from cosmic rays with galactic origin.
It was shown that the GW speed inside the Milky Way must satisfy $\alpha_T \geq - 4 \times 10^{-15}$, otherwise Cherenkov radiation would take energy away from cosmic rays, altering the spectrum of cosmic rays we see on Earth~\cite{Moore:2001bv}.
Estimating $\alpha_T$ using eq.~\eqref{eq:alpha_T} with local dark matter density $0.3~\mathrm{GeV/cm^3}$~\cite{Bovy:2012tw,Pato:2015dua}, we have:
\begin{align}
\label{eq:gw_constraint_Cherenkov}
    -3 \times 10^{46}~\mathrm{eV^{-2}} \lesssim \frac{\xi_2}{m^2}
    ~,
\end{align}
where we used averaged value $\expval{\rho_3}=\rho/3$ in the expression for $\alpha_T$.
This lower bound is stronger than the one obtained from GW170817.
Again, this constraint is shown in figure \ref{fig:gw_speed}.

\section{Summary and discussions}
\label{sec:conclusion}

In this paper, we study the phenomenological implications of nonrelativistic wave VDM that is nonminimally coupled to gravity. From the covariant action in terms of the vector field $X_\mu$, we derive a nonrelativistic EFT by extracting the slowly varying complex field $\psi_\mu$ and integrating out fast oscillating terms. In the low-energy limit, nonminimal couplings manifest themselves as effective self-interactions in the modified SPF equations \eqref{EOM_psi}-\eqref{EOM_H} in a perturbed expanding universe. We demonstrate that the nonminimal interaction could impact the mass-radius relation of ground-state vector solitons, growth of linear perturbations during structure formation and the propagating speed of GWs.

The impact of the nonminimal coupling on the mass-radius relation of vector solitons, as shown in figure \ref{fig:mass_radius_relation}, is similar to that of a quartic self-interaction reported in previous works \cite{Chavanis:2011zi, Chavanis:2011zm, Salehian:2021khb}. However, the key difference between these effective self-interactions for $\psi_i$ is that the nonminimal coupling does not depend on specific polarization states, whereas the quartic self-interaction $(X_\mu X^\mu)^2$ results in $\psi_j \psi_j^* \psi_i$ and $\psi_j \psi_j \psi_i^*$ terms in the SPF equations, which break the degeneracy in the energy of polarized vector solitons \cite{Jain:2021pnk, Zhang:2021xxa}. Effective self-interactions due to the nonminimal coupling could relax current constraints obtained from galactic density profiles, alleviate tensions between halo mass and soliton mass in some galaxies, and open up a new window of VDM solitons \cite{Robles:2018fur, Desjacques:2019zhf, Safarzadeh:2019sre}.

We find that, in the regime $\xi_a^2 \gg m^2\MP^2/\rho$ $(a=1,2)$ where the nonminimal coupling can play a crucial role, both small- and large-scale linear perturbations can be boosted significantly, in contrast to the standard wave DM scenario wherein only perturbations with scales larger than the Jeans length can grow. To retain the success of CDM in large scales, we require $\abs{2\xi_1 + \xi_2}/m^{2}\lesssim 10^{50} \mathrm{eV^{-2}}$. As the universe expands and the DM density decreases, the importance of nonminimal couplings decreases and the standard scenario of perturbation evolution in wave DM is recovered, albeit with a slightly modified Jeans scale \eqref{jeans_scale}. Besides the growth of linear perturbations, the nonminimal coupling could also affect matter power spectrum, baryon acoustic oscillations, CMB spectrum, and BBN \cite{Sankharva:2021spi, Ferreira:2020fam, Lague:2021frh}. We leave the investigation of these topics for future work.

Moving beyond scalar perturbations, we find that the nonminimal coupling governed by $\xi_2$ can change the speed of GWs, with the deviation from the light speed characterized by equation \eqref{eq:alpha_T}. The current best constraints on GW speed come from the lack of observation of gravitational Cherenkov radiation of cosmic rays with a galactic origin \cite{Moore:2001bv}, and the difference in arrival time between GW 170817 \& GRB 170817A \cite{LIGOScientific:2017zic, Baker:2017hug}. These constraints require that the nonminimally coupled VDM satisfies $-3\times 10^{46} \mathrm{eV^{-2}} \lesssim \xi_2 / m^2 \lesssim 8 \times 10^{48} \mathrm{eV^{-2}}$, as illustrated in figure \ref{fig:gw_speed}. Note that the deviation \eqref{eq:alpha_T} is proportional to the partial VDM energy density contributed by the component of the field $\boldsymbol\psi$ in the GW propagating direction, which leads to anisotropies in the GW speed.

\acknowledgments
We would like to thank Dorian Amaral, Mustafa Amin, Ray Hagimoto, Mudit Jain,  Andrew Long, Zong-Gang Mou and Enrico Schiappacasse for helpful comments and discussions. HYZ would like to thank Fengwei Yang (University of Utah) and Jiashu Han (UC San Diego) for discussions on GWs. HYZ is partly supported through a DOE grant DE-SC0021619, and SL is partly supported by the NASA ATP award 80NSSC22K0825. The perturbative calculations for deriving the gravitational wave speed were performed using \texttt{Mathematica} package collection \texttt{xAct}~\cite{Brizuela:2008ra}, which is available for free on the website \url{http://xact.es/faq.html}.

\appendix
\section{Numerical solutions for vector solitons}
\label{app:soliton_shooting}

In this appendix we will outline an approach for obtaining vector soliton profiles numerically. Treating the VDM field equations as wave equations, a massive wave with momentum ${\boldsymbol p}$ can be decomposed into different polarizations
\begin{align}
	\psi_\mu(t,{\boldsymbol x}) = \sum_{\sigma=0,\pm 1} e_\mu({\boldsymbol p}, \sigma) \psi(t,{\boldsymbol x}, \sigma) ~,
\end{align}
where $\sigma=0,\pm 1$ is the spin at $z$-direction and $e_\mu({\boldsymbol p}, \sigma)$ is the polarization vector. In the rest frame of the massive wave, ${\boldsymbol p}=0$ and the polarization vectors are chosen as
\begin{align}
	e^\mu(0,0) =
	\begin{pmatrix}
		0 \\ 0 \\ 0 \\ 1
	\end{pmatrix}~,\quad
	e^\mu(0,1) = - \frac{1}{\sqrt{2}}
	\begin{pmatrix}
		0 \\ 1 \\ i \\ 0
	\end{pmatrix}~,\quad
	e^\mu(0,-1) = \frac{1}{\sqrt{2}}
	\begin{pmatrix}
		0 \\ 1 \\ -i \\ 0
	\end{pmatrix} ~.
\end{align}

The soliton profile can be found by considering ansatz with each polarization
\begin{align}
	\label{soliton_ansatz}
	\psi_i (t,{\boldsymbol x},\sigma) = m^{1/2} \MP f(r,\sigma) e_i(0,\sigma) e^{i\mu t} ~,
\end{align}
where we have assumed a spherically symmetric profile $f(r)$, and $\mu>0$ in order that the vector particles be bounded. The solutions $\psi_i(t,{\boldsymbol x},0)$ are called directional solitons (longitudinally polarized) and $\psi_i(t,{\boldsymbol x},\pm 1)$ are called spinning solitons (circularly polarized) \cite{Jain:2021pnk, Zhang:2021xxa}. Plugging the ansatz \eqref{soliton_ansatz} into the SPF equations \eqref{EOM_psi}-\eqref{EOM_H} with $a(t)=1$, we obtain the following profile equations for all three polarizations $\sigma=0,\pm 1$:
\begin{align}
	\label{soliton_profile_eq1}
	& \nabla^2 f - 2(m\mu+m^2\Phi_N) f - 4m^2 \Phi_\xi f + (2\xi_1 + \xi_2) (\nabla^2\Phi_\xi) f = 0 ~,\\
	\label{soliton_profile_eq2}
	& \nabla^2\Phi_N = \frac{1}{2} m^2 f^2 ~,\quad
	\Phi_\xi = - \frac{1}{4}(2\xi_1 + \xi_2)f^2 ~.
\end{align}
where
\begin{align*}
	\nabla^2f = r^{-2} \partial_r[ r^2 \partial_rf ] = \partial_r^2 f + (2/r)\partial_rf ~,\quad
	\nabla^2 f^2 = r^{-2} \partial_r[ r^2 \partial_r(f^2) ] = 2[ (\partial_rf)^2 + f\nabla^2f ] ~.
\end{align*}
If we replace $\Phi$ by $\Psi\equiv \Phi + \mu/m$, then $\mu$ disappears, and the profile equations \eqref{soliton_profile_eq1} and \eqref{soliton_profile_eq2} can be solved by using the numerical shooting method for each value of the amplitude of $f(r)$ at $r=0$. Once the soliton profiles have been obtained numerically, we can plug them into \eqref{soliton_number_mass} and obtain the numerical values of the coefficients in the mass-radius relation \eqref{eq:MR}.

\section{Derivation of gravitational wave speed}
\label{app:GW_speed}


In this appendix, we will derive the speed $c_T$ of a gravitational wave propagating on a background of nonrelativistic and nonminimally coupled ultralight vector dark matter.
Since the VDM's gravitational potential and the GW are both described by the metric, we write the full metric as $g\indices{_\mu_\nu} = \eta\indices{_\mu_\nu} + \gamma\indices{_\mu_\nu} + h\indices{_\mu_\nu}$, where $\gamma\indices{_\mu_\nu}$ is induced by the VDM, and $h\indices{_\mu_\nu}$ is the perturbation corresponding to the GW.
We assume $| h\indices{_\mu_\nu} | \ll | \gamma\indices{_\mu_\nu} |$, so that the GW do not back-react on the VDM background.
We will call $\bar{g}\indices{_\mu_\nu} \equiv \eta\indices{_\mu_\nu} + \gamma\indices{_\mu_\nu}$ the background metric, since it is the background on which the GW propagates.
The full metric is then given by $g\indices{_\mu_\nu} = \bar{g}\indices{_\mu_\nu} + h\indices{_\mu_\nu}$, and the GW speed will be obtained by computing the dispersion relation of $h\indices{_\mu_\nu}$.

We first discuss the background, given by a non-zero vector field $X_\mu$ and a curved metric $\bar{g}\indices{_\mu_\nu}$, which together constitute a solution to action \eqref{eq:full_action}.
Let $\nabla_\mu$, $R\indices{_\mu_\nu}$, $R$, $G\indices{_\mu_\nu}$, $\Gamma\indices{^\lambda_\mu_\nu}$ be the covariant derivative, the Ricci tensor, the Ricci scalar, the Einstein tensor and the Christoffel symbol associated with $\bar{g}\indices{_\mu_\nu}$, respectively.\footnote{More precisely, we define $\Gamma\indices{^\lambda_\mu_\nu}$ to be the difference between the connections associated with $\bar{g}\indices{_\mu_\nu}$ and $\eta\indices{_\mu_\nu}$, which means $\Gamma\indices{^\lambda_\mu_\nu}$ is in fact a tensor.}
To characterize the smallness of the field $X_\mu$, we write $X_\mu = \order{\epsilon}$, where $\epsilon$ is a formal parameter.\footnote{More formally, we consider a 1-parameter family of solutions $(X_\mu(\epsilon), \bar{g}\indices{_\mu_\nu}(\epsilon))$, such that $(X_\mu(\epsilon), \bar{g}\indices{_\mu_\nu}(\epsilon))$ is a solution to action \eqref{eq:full_action} for each $0 \leq \epsilon \leq 1$, and that $X_\mu(0) = 0$, $\bar{g}\indices{_\mu_\nu}(0) = \eta\indices{_\mu_\nu}$, $X_\mu(1) = X_\mu$ and $\bar{g}\indices{_\mu_\nu}(1) = \bar{g}\indices{_\mu_\nu}$. See section 7.5 of ref.~\cite{Wald:1984rg} for more details. }
Since the stress-energy tensor $T\indices{_\mu_\nu} \equiv (-2 / \sqrt{-g}) \fdv*{S_M}{g\indices{^\mu^\nu}}$ is quadratic in $X_\mu$ and its derivatives, we have $T\indices{_\mu_\nu} = \order{\epsilon^2}$.
The Einstein equation $G\indices{_\mu_\nu} = \MP^{-2} T\indices{_\mu_\nu}$ then implies $R, R\indices{_\mu_\nu}, G\indices{_\mu_\nu} = \order{\epsilon^2}$.
The formulas for $R\indices{_\mu_\nu}$ in terms of $\Gamma\indices{^\lambda_\mu_\nu}$ and $\gamma\indices{_\mu_\nu}$ imply
that are of same order in $\epsilon$,
so $\Gamma\indices{^\lambda_\mu_\nu}$ and $\gamma\indices{_\mu_\nu}$ are also $\order{\epsilon^2}$.
In summary:
\begin{align}
    \label{eq:metric_perturbation_orders}
    \gamma\indices{_\mu_\nu}, \Gamma\indices{^\lambda_\mu_\nu}, R, R\indices{_\mu_\nu}, G\indices{_\mu_\nu} = \order{\epsilon^2}
    ~.
\end{align}
The above information will be used repeatedly for the rest of this appendix.

We now show that $X_\mu$ evolves like a minimally coupled Proca field in the linear regime.
Taking variation of action \eqref{eq:full_action} with respect to $X_\nu$, we obtain the EOM for $X^\nu$:
\begin{align*}
  0 &= \nabla_\mu \nabla^\mu X^\nu - m^2 X^\nu + \Big[ \xi_1 R X^\nu + \xi_2 R\indices{^\nu_\mu} X^\mu - ( \nabla^\nu \nabla_\mu X^\mu + R\indices{_\rho^\nu} X^\rho ) \Big]
    \;.
\end{align*}
Taking the divergence of the EOM gives us a constraint:
\begin{align*}
  \nabla_\nu X^\nu &= \frac{1}{m^2 - \xi_1 R} \Big[ \xi_1 X^\nu \nabla_\nu R + \xi_2 (R\indices{_\nu_\mu} \nabla^\mu X^\nu + X^\nu \nabla_\mu R\indices{_\nu^\mu}) \Big]
                     \;.
\end{align*}
By \eqref{eq:metric_perturbation_orders}, the RHS above is $\order{\epsilon^3}$, so $\nabla_\nu X^\nu = \order{\epsilon^3}$. 
Eliminating $\nabla^\nu (\nabla_\mu X^\mu)$ from the EOM then yields $\nabla_\mu \nabla^\mu X^\nu - m^2 X^\nu = \order{\epsilon^3}$.
In summary:
\begin{align}
  \label{eq:approximate_Proca_eqns}
    \nabla_\mu \nabla^\mu X^\nu - m^2 X^\nu = \order{\epsilon^3} \qq{and}
    \nabla_\mu X^\mu = \order{\epsilon^3}
    \;.
\end{align}
The above results can be interpreted as: in the linear regime,
the EOM and the constraint for $X_\mu$ reduce to that for a minimally coupled Proca field.

We now proceed to find the EOM for the gravitational wave $h\indices{_\mu_\nu}$.
Expanding the Lagrangian in eq.~\eqref{eq:full_action} around the background $\bar{g}\indices{_\mu_\nu}$ to quadratic order in $h\indices{_\mu_\nu}$ gives us 
\begin{align}
  \mathcal{L}^{\textrm{(GW)}} &= \mathcal{L}_G^{\textrm{(GW)}} + \mathcal{L}_M^{\textrm{(GW)}} \nonumber \\
  \mathcal{L}_G^{\textrm{(GW)}} &= \sqrt{-\bar{g}}\, \frac{\MP^2}{4} \Big[ -\frac{1}{2} \nabla\indices{_\lambda} h\indices{_\mu_\nu} \nabla\indices{^\lambda} h\indices{^\mu^\nu}
  + \nabla\indices{_\mu} h\indices{^\nu^\lambda} \nabla\indices{_\nu} h\indices{^\mu_\lambda}
  - \nabla\indices{_\mu} h\indices{^\mu^\nu} \nabla\indices{_\nu} h
  + \frac{1}{2} \nabla\indices{_\mu} h \nabla\indices{^\mu} h \Big] 
  \nonumber \\ & \quad
  + \textrm{($h^2$ terms)} \nonumber \\
  \mathcal{L}_M^{\textrm{(GW)}} &= \sqrt{-\bar{g}}\, \xi_1 X^\lambda X^\rho \Big[
  - \nabla_\sigma h \nabla^\sigma h\indices{_\lambda_\rho}
  + \nabla_\sigma h\indices{_\lambda_\rho} \nabla_\mu h\indices{^\sigma^\mu}
  \nonumber \\ & \qquad
  + g\indices{_\lambda_\rho} \Big( \frac14 \nabla_\sigma h \nabla^\sigma h
  - \frac12 \nabla_\sigma h \nabla_\mu h\indices{^\sigma^\mu}
  + \frac12 \nabla_\sigma h\indices{_\rho_\mu} \nabla^\mu h\indices{^\rho^\sigma}
  - \frac14 \nabla_\sigma h\indices{_\mu_\nu} \nabla^\sigma h\indices{^\mu^\nu} \Big)
  \Big]
  \nonumber \\ & \quad
  + \sqrt{-\bar{g}}\, \xi_2 X^\lambda X^\rho \Big[
  - \frac14 \nabla_\lambda h\indices{_\mu_\nu} \nabla_\rho h\indices{^\mu^\nu}
  + \frac14 \nabla_\lambda h \nabla_\rho h
  - \nabla_\rho h \nabla^\sigma h\indices{_\lambda_\sigma}
  \nonumber \\ & \qquad
  + \nabla_\rho h\indices{_\mu_\nu} \nabla^\nu h\indices{_\lambda^\mu}
  + \frac12 \nabla_\mu h\indices{_\rho_\nu} \nabla^\nu h\indices{_\lambda^\mu}
  - \frac12 \nabla_\nu h\indices{_\rho_\mu} \nabla^\nu h\indices{_\lambda^\mu}
  \Big]
  \nonumber \\ & \quad
  + \textrm{($h^2$ terms)} + \order{\epsilon^3}
  ~,
\end{align}
where $h \equiv \bar{g}\indices{^\mu^\nu} h\indices{_\mu_\nu}$ is the trace.
To further simplify the Lagrangian, we choose the transverse traceless gauge whereby:
\begin{align}
    \label{eq:tt_gauge}
    \nabla^\mu h\indices{_\mu_\nu} = 0,\quad
    h = 0
    ~.
\end{align}
The Lagrangian simplifies under this gauge choice to:
\begin{align}
  \mathcal{L}^{\textrm{(GW)}} &= 
                                \sqrt{-\bar{g}}\, \Big[
                                -\frac18 \MP^2 \nabla\indices{_\lambda} h\indices{_\mu_\nu} \nabla\indices{^\lambda} h\indices{^\mu^\nu}
                                - \frac18 \xi_1 X_\lambda X^\lambda \nabla_\sigma h\indices{_\mu_\nu} \nabla^\sigma h\indices{^\mu^\nu}
                                \nonumber \\ & \quad
                                - \frac18 \xi_2 X^\lambda X^\rho \left( \nabla_\lambda h\indices{_\mu_\nu} \nabla_\rho h\indices{^\mu^\nu}
                                + 2 \nabla_\nu h\indices{_\rho_\mu} \nabla^\nu h\indices{_\lambda^\mu} \right)
                                \Big]
                                \nonumber \\ & \quad + \textrm{($h^2$ terms)} + \order{\epsilon^3} 
  ~.
\end{align}
Taking variation of above with respect to $h\indices{^\mu^\nu}$, we obtain the EOM for $h\indices{_\mu_\nu}$:
\begin{align}
  \label{eq:h_eom_covariant}
  0=& \left[1 + \frac{\xi_1 X_\rho X^\rho}{\MP^2} \right] \nabla_\lambda \nabla^\lambda h\indices{_\mu_\nu}
      + \left[ \frac{\xi_2 X^\rho X_\nu}{\MP^2} \nabla_\lambda \nabla^\lambda h\indices{_\mu_\rho}
      + \frac{\xi_2 X_\mu X^\rho}{\MP^2} \nabla_\lambda \nabla^\lambda h\indices{_\rho_\nu} \right]
      \nonumber \\ &
                     + \frac{\xi_2 X^\lambda X^\rho}{\MP^2} \nabla_\rho \nabla_\lambda h\indices{_\mu_\nu}
                     + \textrm{($\nabla h$ and $h$ terms)} + \order{\epsilon^3}
                     \;.
\end{align}
The $\xi_1$ and $\xi_2$ terms above are $\order{\epsilon^2}$; they give the leading order correction to the dispersion relation for $h\indices{_\mu_\nu}$.
Note that we have neglected $\nabla h\indices{_\mu_\nu}$, $h\indices{_\mu_\nu}$ and $\order{\epsilon^3}$ terms in the EOM,
since they do not modify the dispersion relation at leading order.

Our discussion so far has been coordinate-free.
To proceed, we derive the coordinate expression for a nonrelativistic field $X_\mu$.
Since spatial gradients of a nonrelativistic field are small, locally we may take a spatially coherent ansatz $X_\mu = \Re[ (Y_t(t), \vb*{Y}(t)) ]$, where $Y_t$ and $\vb*{Y}$ take complex values.
Under this ansatz, equations \eqref{eq:approximate_Proca_eqns} imply:
\begin{align}
  & - \partial_t^2 Y_t - m^2 Y_t = \order{\epsilon^3},\quad
  - \partial_t Y_t = \order{\epsilon^3},\quad \nonumber \\
  & - \partial_t^2 \vb*{Y} - m^2 \vb*{Y} = \order{\epsilon^3}
    \;,
\end{align}
where we have traded covariant derivatives for coordinate derivatives.
Solving the equations gives us:
\begin{align}
    \vb*{Y}(t) = A e^{i m t} + B e^{- i m t} + \order{\epsilon^3} \qq{and} Y_t = \order{\epsilon^3}
    \;.
\end{align}
Since $X_\mu = \order{\epsilon}$, the above results show that $\vb*{Y} = \order{\epsilon}$, and that the time evolution of $X_\mu$ is dominated by oscillation of frequency $m$.
Finally, we conclude without loss of generality that:
\begin{align}
\label{eq:x_mu_approx}
  X_\mu(t,\vb{x})
  = (0, \vb*{X}) + \order{\epsilon^3}
  = (0, \Re[ e^{i m t} \vb*{Y} ]) + \order{\epsilon^3}
  \;.
\end{align}
where $\vb*{Y}$ is now a constant complex vector,
and $\vb*{X} = \Re[ e^{imt} \vb*{Y} ]$.

We will now derive the wave equations for the individual polarizations of the GW.
The two polarization tensors for the GW are given by:
\begin{align}
\label{eq:h_polarization_tensor}
    E_+ = \sqrt{2} \mqty(
    0 & 0 & 0 & 0 \\
    0 & 1 & 0 & 0\\
    0 & 0 & -1 & 0\\
    0 & 0 & 0 & 0),\quad
    E_\times = \sqrt{2} \mqty(
    0 & 0 & 0 & 0 \\
    0 & 0 & 1 & 0\\
    0 & 1 & 0 & 0\\
    0 & 0 & 0 & 0)
    ~.
\end{align}
These polarization tensors are approximations of the exact ones satisfying the gauge condition \eqref{eq:tt_gauge}.
The GW is then parameterized as:
\begin{align}
  \label{eq:h_mode_decomposition}
  h\indices{_\mu_\nu} = 
  h_+(t,z) [E_+]\indices{_\mu_\nu} 
  + h_\times(t,z) [E_\times]\indices{_\mu_\nu}
  + \order{\epsilon^2}
  ~.
\end{align}
We plug the above ansatz and the coherent $X_\mu$ background \eqref{eq:x_mu_approx} into the $h\indices{_\mu_\nu}$ EOM \eqref{eq:h_eom_covariant}.
This yields the EOM for $h_+$ and $h_\times$:
\footnote{There is an alternate way to calculate GW speed: we assume the metric background is Minkowski, expand $g\indices{_\mu_\nu} = \eta\indices{_\mu_\nu} + h\indices{_\mu_\nu}$, and find the correction to the GW dispersion relation by expanding the nonminimal terms to quadratic order in $h\indices{_\mu_\nu}$ on the given $X_\mu$ background.
  The reason why this alternate calculation works is as follows.
  If we write the wave equation \eqref{eq:h_eom_covariant} in terms of coordinate derivatives,
  the extra terms involving Christoffel symbols do not contain double derivatives of $h\indices{_\mu_\nu}$.
  These terms describe effects such as gravitational redshift and lensing, and they do not affect the dispersion relation.
  We may thus perform the calculation as if the Christoffel symbols are zero,
  which amounts to replacing $\bar{g}\indices{_\mu_\nu}$ with $\eta\indices{_\mu_\nu}$.
}
\begin{align}
  \label{eq:GW_EOM}
  0=& \left[1 + \frac{\xi_1}{\MP^2} |\vb*{X}|^2 + \frac{\xi_2}{\MP^2} (|\vb*{X}|^2 - |X_z|^2) \right] \left[ \partial_t^2 h_{\lambda} - \partial_z^2 h_{\lambda} \right]
      - \frac{\xi_2 |X_z|^2}{\MP^2} \partial_z^2 h_{\lambda}
      \nonumber \\ &
                     + \textrm{($h$ and $\partial h$ terms)} + \order{\epsilon^3} \qq{for} \lambda = +, \times
                     \;.
\end{align}

There are several features in the $h_\lambda$ EOM \eqref{eq:GW_EOM}.
Firstly, writing the EOM in the form of the wave equation $0 = \partial_t^2 h_{\lambda} - c_T^2 \partial_z^2 h_{\lambda} + \hdots$ gives us the correction to the GW speed:
\begin{align}
\label{eq:alpha_T_X}
    \alpha_T \equiv c_T^2 - 1
    = \frac{\xi_2 |X_z|^2}{\MP^2} + \order{\epsilon^3}
    = \frac{\xi_2 |Y_z|^2 \cos^2(mt + \arg Y_z)}{\MP^2} + \order{\epsilon^3}
    ~.
\end{align}
One can see that only the $z$-component (the direction of GW propagation) of $\vb*{X}$ contributes to the GW speed.
Moreover, $\alpha_T$ is oscillatory in time with frequency $2m$, and the magnitude of $\alpha_T$ is controlled by the constant $|Y_z|$.
Secondly, the EOM for the two polarizations of the GW are the same, so there is no birefringence of GW, at least for the plane wave ansatz that we consider.
Thirdly, the $h$ and $\partial h$ terms lead to effects such as gravitational redshift and lensing, and they can be ignored for GW speed discussions.


Finally, we employ the nonrelativistic expansion \eqref{eq:nr_expansion} to write $\alpha_T$ in terms of the mass density component $\rho_3$.
Ignoring the expansion of the universe by setting $a(t)=1$, the nonrelativistic field corresponding to $X_\mu(t, \vb{x}) = (0, \Re[e^{imt} \vb*{Y}])$ is $\psi_\mu(t,\vb{x}) = \sqrt{m/2}\, (0, \vb*{Y}^\ast)$, and the time-averaged speed correction is:
\begin{align}
\label{eq:alpha_T_psi}
    \expval{\alpha_T}
    \approx 
    \frac{\xi_2 |\psi_3|^2}{m \MP^2}
    = \frac{\xi_2 \rho_3}{m^2 \MP^2}
    ~.
\end{align}

\bibliographystyle{jhep}
\bibliography{main}
\end{document}